\documentclass[aps,prd,letterpaper,11pt,twoside,tightenlines,nofootinbib,showpacs,preprint,twocolumn]{revtex4}
\usepackage{graphicx}
\usepackage[sort&compress]{natbib}
\usepackage{latexsym}
\usepackage{epsfig}
\usepackage[english]{babel}
\usepackage{graphicx}
\usepackage[T1]{fontenc}
\usepackage[utf8]{inputenc}
\usepackage{amsmath}

\usepackage[figurename=FIG.,tablename=TABLE,font=footnotesize]{caption}

\usepackage{bm}

\begin{document}
\newcommand{\eg}{{\it e.g.}}
\newcommand{\etal}{{\it et. al.}}
\newcommand{\ie}{{\it i.e.}}
\newcommand{\be}{\begin{equation}}
\newcommand{\dd}{\displaystyle}
\newcommand{\ee}{\end{equation}}
\newcommand{\bea}{\begin{eqnarray}}
\newcommand{\eea}{\end{eqnarray}}
\newcommand{\bef}{\begin{figure}}
\newcommand{\eef}{\end{figure}}
\newcommand{\bce}{\begin{center}}
\newcommand{\ece}{\end{center}}

\title{Deconfinement Transition Effects on  Cosmological Parameters and Primordial Gravitational Wave Spectrum }
\author{P. Castorina$^{1,2}$, D. Lanteri$^{1,2}$, S. Mancani$^1$}
\affiliation{
\mbox{${}^1$ Dipartimento di Fisica, Universit\`a di Catania, Via Santa Sofia 64,
I-95123 Catania, Italy.}\\
\mbox{${}^2$ INFN, Sezione di Catania, I-95123 Catania, Italy.}\\
}

\date{\today}
\begin{abstract}

The cosmological evolution can be described  in terms of directly measurable cosmological scalar parameters (deceleration $q$, jerk $j$, snap $s$, etc...)  constructed out of high order derivatives of the scale factor. Their behavior at the critical temperature of the Quantum Chromodynamics (QCD) phase transition in early universe could be a specific tool to study  the transition,  analogously to the fluctuations of conserved charges in QCD. We analyze the effect of the crossover transition from quarks and gluons to hadrons in early universe on the cosmological scalars and on the gravitational wave spectrum, by using the  recent lattice QCD equation of state and including  the electroweak degrees of freedom and different models of dark matter. Near the transition the cosmological parameters follow the behavior of QCD trace anomaly and of the speed of sound of the entire system. The effects of deconfinement turn out to be more relevant for the modification of the primordial spectrum of gravitational waves  and our complete analysis, based on lattice QCD simulations and on the hadron resonance gas below the critical temperature, refines previous results.

\end{abstract}
 \maketitle

\section*{Introduction}

Quantum Chromodynamics (QCD) deconfinement phase transition has a interesting role at cosmological level, modifying, for example, the primordial spectrum of the gravitational waves~\cite{Schwarz2003,Schwarz1998,Watanabe2006}.

Other consequences of the  QCD transition show up in the cosmological parameters (deceleration $q$, jerk $j$, etc.) which involve the derivatives of the scale factor $a(t)$~\cite{Castorina2015}. Indeed,  the fluctuations of the cosmological parameters  with higher order derivatives are strongly enhanced by the phase transition.

This effect is similar to the fluctuations of conserved charges (net baryon-number, net electric charge, net strangeness)  evaluated in lattice QCD at finite temperature, which  require the calculation of the higher order cumulants, i.e.  high order derivatives of the  logarithm of the QCD partition function. 
These fluctuations provide a wealth of information on the properties of strong-interaction matter in the transition region from the low temperature hadronic phase to the quark-gluon plasma phase and, in particular, they can be used to quantify deviations from the hadron resonance gas (HRG) model~\cite{Karsch2017,Karsch2003}.

Previous analyses~\cite{Castorina2015,Greco2014,Florkowski2011} considered the evolution of the first cosmological parameters ($q$, $j$) and of the energy density fluctuations during the deconfinement transition  by a specific parametrization of the QCD equation of state (EoS) and  neglecting the dark matter contribution. 

In this paper we discuss the behavior of a larger set of cosmological scalars, with higher order derivatives of the scale factor, and  take into account the electroweak sector, the strongly interacting sector and different models of dark matter also. Moreover the transition (cross-over) between the quark-gluon phase  and the hadronic phase is described by recent lattice QCD EoS~\cite{HotQCD2014}  and by the HRG~\cite{Huovinen2010} below the critical temperature $T_c \simeq 150$ MeV.

Finally, the detailed treatment of the EoS above and below $T_c$ permits a refined analysis of the modification of the primordial spectrum of the gravitational waves. 

The paper is organized as follows: the definition of the cosmological parameters is given in Sec.~\ref{sec:1}; the relevant degrees of freedom and the role of the different contributions to the total energy density and to the EoS of the whole system are discussed in Sec.~\ref{sec:2}; Sections~\ref{sec:3} and \ref{sec:4} contain respectively  the results on the speed of sound and on the  fluctuations of the cosmological parameters during the deconfinement transition, with different dark matter models; Sec.~\ref{sec:5} is devoted to the modification of the primordial gravitational wave spectrum due to the transition; comments and conclusions are in Sec.~\ref{sec:6}.

\section{Cosmological parameters\label{sec:1}}
The standard model of the cosmological evolution is based on the Friedmann - Lemaître - Robertson - Walker (FLRW) equations and a set of  equations of state for the different contributions to the total energy density. 
By defining the total energy density $\varepsilon_T$ and the total pressure $p_{T}$ as 
\begin{equation}\label{eq:Eeff}
\varepsilon_T = \varepsilon_{s} + \varepsilon_{ew} + \varepsilon_{d} + \varepsilon_{\Lambda}\;,
\end{equation}
\begin{equation}
p_T = p_{s} + p_{ew} + p_{d} + p_{\Lambda}\;,
\end{equation}
where 
\begin{equation}
\varepsilon_\Lambda \equiv \frac{\Lambda}{8\,\pi\,G}\;,
\ee
\be
p_\Lambda \equiv - \varepsilon_\Lambda\;,
\end{equation}
are the dark energy contributions and the other terms correspond to strong (s), electroweak (ew) and dark matter (d) sector,
the FLRW equations for a flat Universe are given by
\begin{equation}\label{eq:FRWL}
\begin{split}
\left(\frac{1}{a}\frac{da}{dt}\right)^2 = & \frac{8\,\pi\,G}{3}\;\varepsilon_T \;,
\\
\frac{1}{a}\frac{d^2 a}{dt^2}
= &
-\frac{4\,\pi\,G}{3}\left(\varepsilon_T+3\,p_T\right)\;.
\end{split}
\end{equation}
The cosmological parameters are defined as~\cite{Visser2004,Dunajski2008}
\begin{equation}\label{eq:cp}
\begin{split}
H\equiv & \frac{1}{a}\frac{da}{dt} \;,
\qquad 
q \equiv -\frac{1}{a\,H^2}\;\frac{d^2a}{dt^2} \; ,\\
A_n \equiv & \frac{1}{a\,H^n}\;\frac{d^na}{dt^n} \qquad (n>2)
\end{split}
\end{equation} 
and their evolution is directly related to the EoS. Indeed, $A_n$ can be written as the sum of terms containing the first $n-1$ derivatives of the Hubble parameter $H$, which can be expressed in terms of the $w\equiv p_T/\varepsilon_T$, of the  speed of sound, $c^2_s\equiv \partial p_T/ \partial \varepsilon_T$, and  its derivatives.
For example, the jerk, $j$, is given by
\begin{equation}
j = A_3 = 1 + 3\,\frac{\dot H}{H^2} + \frac{\ddot H}{H^3}
\end{equation}
and, by FLRW equations, one has
\begin{equation}
\begin{split}
\frac{\dot H}{H^2} = & -\frac{3}{2} \left(1 + \frac{p_T}{\varepsilon_T}\right) \; ,\\
\frac{\ddot H}{H^3} = & \frac{9}{2} \left(1 + c^2_s\right) \left(1 + \frac{p_T}{\varepsilon_T}\right)\;.
\end{split}
\end{equation}
The complete set of relations for various cosmological parameters is given in Appendix~\ref{app:a}.

The cosmological evolution can be described by the Hubble parameter $H$, the deceleration $q$, the jerk $j$, the snap ($s=A_4$) and the others cosmological parameters since they specify the various terms of the Taylor expansion of the scale factor:
\begin{equation}
\begin{split}
a(t) \!= \!a(t^*)\Big[ & 1 + H(t^*)\left(t-t^*\right)
- \\
& -
\frac{\left(q\,H^2\right)(t^*)}{2!}\left(t-t^*\right)^2 +\\
&
+
\frac{\left(j\,H^3\right)(t^*)}{3!}\left(t-t^*\right)^3+\cdots
\Big]\;.
\end{split}
\end{equation}
In Sec.~\ref{sec:4} the effect of the QCD deconfinement transition on the cosmological parameters will be analyzed and, as discussed in the introduction, higher order derivatives of $a(t)$  show larger fluctuations.

\section{The equation of state in the early Universe\label{sec:2}} 

Early Universe was a hot and dense plasma and during the cosmological evolution the number of degrees of freedom changed due to  various phase transitions (see Fig. 1.1 of ref.~\cite{Schwarz2003}). Since we are interested in the effect of the deconfinement phase transition on the cosmological parameters and on the spectrum of gravitational waves,  we consider the temperature $T$ in the range  $ 70$ MeV $ < T < 400 $ MeV. In this section the number of degrees of freedom and the equations of state for strong, electroweak and dark matter sectors will be discussed.

\subsection{a. Strong and electroweak sectors}

The QCD deconfinement transition rapidly reduces the number of the strongly interacting degrees of freedom, $g_s$. However, lattice QCD simulations indicates that the transition is not so sharp and is indeed a cross-over between a system of quarks and gluons and a hadron gas~\cite{HotQCD2014,Borsanyi2014,Borsanyi2010}. 
The (pseudo) critical temperature turns out to be $T_c \sim 150-160\,MeV$ by the analysis of chiral susceptibility.

Starting from the lattice QCD partition function,  one  defines the trace anomaly $\Theta^{\mu\mu}(T)$  as the derivative with respect to the lattice spacing $a_l$~\cite{HotQCD2014}
\begin{equation}
\Theta^{\mu\mu}(T) = - \frac{T}{V}\;\frac{d \ln \mathcal Z}{d \ln a_l}  
\end{equation}
and one evaluates all other thermodynamical quantities, i.e. the pressure
\begin{equation}\label{eq:p}
\frac{p(t)}{T^4} = \frac{p_0}{T_0^4} + \int_{T_0}^T\,dT^\prime \,\frac{\Theta^{\mu\nu}(T^\prime)}{T^{\prime^{\scriptstyle 5}}}\; ,
\end{equation}
the energy density 
\begin{equation}
\varepsilon = 3\,p+\Theta^{\mu\nu}\;,
\end{equation}
the entropy density $s$
\begin{equation}\label{eq:entropy}
s = \frac{\varepsilon + p}{T}\;, 
\end{equation}
and the speed of sound
\begin{equation}
c^2_s = \frac{\partial p}{\partial \varepsilon} 
=\frac{s}{C_V}= \frac{\partial p/\partial T}{\partial \varepsilon/\partial T}\;,  
\end{equation}
where $C_V$ is the specific heat.

The pressure obtained by the HotQCD collaboration can be parametrized as follows~\cite{HotQCD2014}:
\begin{equation}\label{eq:plattice}
p^{lattice}(T) = \frac{T^4}{2}\,\left[1 + \tanh\left[c_t\left(t - t_0\right)\right]\right]f(T),
\end{equation}
where
\begin{equation}\label{eq:plattice2}
f(T)=\frac{{p_{id} + \frac{a_n}{t} +\frac{b_n}{t^2}+\frac{c_n}{t^3}+\frac{b_n}{t^4}} }{{1+
		\frac{a_d}{t} +	\frac{b_d}{t^2}+\frac{c_d}{t^3}+\frac{b_d}{t^4}}}
\end{equation}
and  $t = T/T_c$, $T_c=154\,MeV$, $p_{id} = 95/180 \pi^2$ is the ideal gas value of $p/T^4$ for massless 3-flavor QCD and the other parameters are summarized in Table~\ref{tab:lattice}.

\begin{table}[t]
	\begin{center}
		\begin{tabular}{|c|c|c|c|c|}
			\hline
			$\quad\;\;c_t\quad\;\;$&$\quad\;\;a_n\quad\;\;$&$\quad\;\;b_n\quad\;\;$&$\quad\;\; c_n \quad\;\; $&$ \quad\;\; d_n\quad\;\;$ \\
			3.8706&-8.7704&3.9200&0&0.3419 \\
			\hline
			\hline
			$t_0$&$a_d$&$b_d$&$c_d$&$d_d$ \\
			0.9761&-1.2600&0.8425&0&-0.0475 \\
			\hline
		\end{tabular}
	\end{center}
	\caption{Parameters used  in Eqs.~\eqref{eq:plattice} and \eqref{eq:plattice2} for the pressure of (2+1)-flavor QCD in the temperature interval
		$T\in [100~{\rm MeV}, 400~{\rm MeV}]$~\cite{HotQCD2014}.}
	\label{tab:lattice}
\end{table}

In the temperature region $T<T_c$  all thermodynamic quantities are well described by the hadron resonance gas (HRG) model where the gran canonical partition function can be expressed as a sum of  one-particle partition functions $\mathcal Z^1_i$ over all hadrons and resonances~\cite{Venugopalan1992}.  
If $m_{max}$ is the maximum mass one includes,  the trace anomaly can be written as a sum over all particles species with mass $m_i\leq m_{max}$~\cite{HotQCD2014,Karsch2003},
\begin{equation}
\begin{split}
\left(\frac{\Theta^{\mu\mu}}{T^4}\right)^{HRG}
&=
 \sum_{m_i\leq m_{max}}\frac{d_i}{2\,\pi^2} \times
\\
& \times \sum_{k=1}^{\infty} 
\frac{(-\eta_i)^{k+1}}{k}(\frac{m_i}{T})^3K_1(\frac{k\,m_i}{T})\;,
\end{split}
\end{equation}
where $\eta_i=-1(+1)$ for bosons (fermions), $K_1$ is the modified Bessel function, $d_i$ are the degeneracy factors.

The trace anomaly for the HRG has been parametrized as~\cite{Huovinen2010} 
\begin{equation}\label{eq:tHRG}
\left(\frac{\varepsilon - 3\,P}{T^4}\right)^{HRG} \!\!\!\!= a_1 T + a_2 T^3+a_3 T^4+a_4 T^{10}\;,
\end{equation}
with $a_i$ given in Table~\ref{tab:HRG} and, by interpolation with lattice data, the pressure is given by
\begin{equation}
p^{HRG}(T) = p_l^{lattice}\left(\frac{T}{T_l}\right)^4
+g(T) \; ,
\end{equation}
where $p_l^{lattice}$ is the pressure at $T_l=130\;MeV$ and 
\begin{equation}\label{eq:tHRG2}
\begin{split}
g(t)=T^4 \big[&a_1(T-T_l) +  \frac{a_2}{3}(T^3-T^3_l)+ \\
&+ \frac{a_3}{4} (T^4-T^4_l)
+\frac{a_4}{10}(T^{10}-T^{10}_l)\big]\;. 
\end{split}
\end{equation}

\begin{table}[t]
	\begin{center}
		\begin{tabular}{|c|c|}
			\hline
			$\qquad\qquad a_1\qquad\qquad$ & $\qquad\qquad a_2\qquad\qquad$ \\
			$4.654\;GeV^{-1}$ & $-879\;GeV^{-3}$ \\
			\hline
			\hline
			$\qquad\qquad a_3 \qquad\qquad$ & $ \qquad\qquad a_4 \qquad \qquad$\\
			$8081\;GeV^{-4}$ & $-7039000\;GeV^{-10}$\\
			\hline
		\end{tabular}
	\end{center}
	\caption{Parameters used in  Eqs.~\eqref{eq:tHRG} and \eqref{eq:tHRG2}~\cite{Huovinen2010}.}
	\label{tab:HRG}
\end{table}

The electroweak sector is included as a relativistic gas of massless particles, i.e.  
\begin{equation}
\varepsilon^{ew} = 3\;p^{ew} = 
g_{ew}\;\frac{\pi^2}{30}\;T^4 \; ,
\end{equation}
where $g_{ew} = 14.45$ is the effective number of electroweak degrees of freedom~\cite{Castorina2015}.

\subsection{b. Dark matter sector}
In previous analyses of the evolution of the cosmological parameters the dark matter contribution has been neglected and we now evaluate its role in the EoS of the complete system by considering  the following different models  (see Appendix~\ref{app:b} for details): 
\begin{itemize}
	\item \textbf{\small{Cold Dark Matter:}}  for pressure, $p_{cdm}$, and energy density, $\varepsilon_{cdm}$,  of the  cold dark matter one has
	\begin{equation}
	\qquad\quad\;
	p_{cdm} = 0 \; ,
	\qquad\qquad
	\varepsilon_{cdm} = \frac{m_{cdm}}{a^3} \;.
	\end{equation}    
	By considering the isentropic condition $a^3\,s=a_0^3\,s_0$, and indicating with $\overline \epsilon$ and $\overline p$ the non-dark-matter contributions to the energy density and to the pressure, one obtains (see Appendix~\ref{app:b1})
	\begin{equation}
		\varepsilon_{cdm} = \frac{\overline \varepsilon + \overline p}{K\,T-1}   \; ,
	\end{equation} 
	where
	\begin{equation}
	\;\;\;\,\,\quad\quad K = \frac{1}{T_0}\left( \frac{\Omega_{b0}}{\Omega_{cdm0}}+1 \right)  \sim \frac{5\times 10^9}{MeV}\;,
	\end{equation}
	$T_0$ is the present temperature, $\Omega_{b0}$ and $\Omega_{cdm0}$ the present baryon and cold dark matter density respectively. 
	
	Since we study the cosmological evolution for temperatures such that $K\,T\gg 1$,
	\begin{equation}
	\varepsilon_{cdm}\simeq \frac{\overline s}{K} \sim 2 \times 10^{-10}\;\overline \varepsilon  
	\end{equation}
	and the cold dark matter term can be neglected.
	
	\item \textbf{\small{General Barotropic case:}}
	a barotropic equation of state is a linear relation between energy density, $\epsilon_b$, and pressure, $p_b$, 
	\begin{equation}
	p_b = w_b \; \varepsilon_b \; ,
	\end{equation}
	with $w_b=const$. Using again the isentropic evolution, one gets, as in the previous case, an equation connecting  the energy of the dark sector to the energy of all other sectors. This equation can be  analytically solved  for a few values of $w_b$  ($w_b=0$ gives the cold dark matter)  and, in general,  one needs numerical methods. We study the value $w_b=1/3$, i.e. a massless dark matter,  useful to estimate the  maximum contribution of the dark sector, although excluded by phenomenological analyses~\cite{chi3}.
	
	\item \textbf{\small{Polytropic case and Chaplygin gas:}} for a polytropic gas  the pressure, $p_p$, and the energy density, $\epsilon_p$, are related by  
	\begin{equation}\label{eq:pp}
	p_p \propto \epsilon_p^\gamma  \; ,
	\end{equation}
	where $\gamma$ is the polytropic index. In Appendix~\ref{app:b3} one shows that, in a cosmological contest, such a case is equivalent to a generalized Chaplygin gas with index $\alpha=-\gamma$, i.e. the energy density is given by
	\begin{equation}\label{eq:ep} 
	\quad\quad\;\;\,\,\,
	\varepsilon_{p} = \varepsilon_p^0\,\left[
	-w_p^0 + \frac{1+w_p^0}{\left(\displaystyle{\frac{a}{a_0}}\right) ^{3(1-\gamma)}} \right]^{\textstyle{\frac{1}{1-\gamma}}} \; ,
	\end{equation}  
	with $w_p^0 \equiv p_p^0/\varepsilon_p^0$ and $a_0$, $p_p^0$ and $\varepsilon_p^0$ evaluated at some reference time. In particular, by choosing $a_0$ at the present time,   
	in the early universe $(a\ll a_0)$ one has to study three different cases:\\
	a) $\gamma=1$ (see Eq.~\eqref{eq:pp}), which is the general barotropic case;\\
	2)  $\gamma<1$, then
	\begin{equation}
    \varepsilon_{p} \simeq \frac{A}{a^3} 
	\end{equation}
	and, by Eqs.~\eqref{eq:entropy} and \eqref{eq:pp}, the entropy density turns out to be
	\begin{equation}
	\qquad\qquad
	T\,s_p \simeq \frac{B}{a^3}\left(1 + C\;a^{3(1-\gamma)} 
	\right) \sim \frac{B}{a^3}\; ,
	\end{equation}
	where  $A$, $B$ and $C$ constants, i.e. with the same behavior of the cold dark matter; \\	
	 3) $\gamma >1$ is physically meaningful only for $w_p^0\leq 0$ and in such a case (see Eq.~\eqref{eq:ep})
	\begin{equation}
	\varepsilon_{p} \propto \left(
	-w_p^0 \right)^{\textstyle{\frac{1}{1-\gamma}}},
	\end{equation}
	\begin{equation}
	p_p = -\varepsilon_p 
	\end{equation}
	and  the resulting entropy density is constant and negligible.

\end{itemize} 

\section{The cosmological deconfinement transition\label{sec:3}}

By previous parametrization of the EoS of strongly interacting, electroweak  and dark matter sectors, we now analyze the  full Eos in the range $70$~MeV~$\leq T\leq 400$~MeV  by interpolating the lattice data and the HRG results at  $ T_l \simeq 130$~MeV.     

\begin{figure}
	\centering
	\includegraphics[width=\columnwidth]{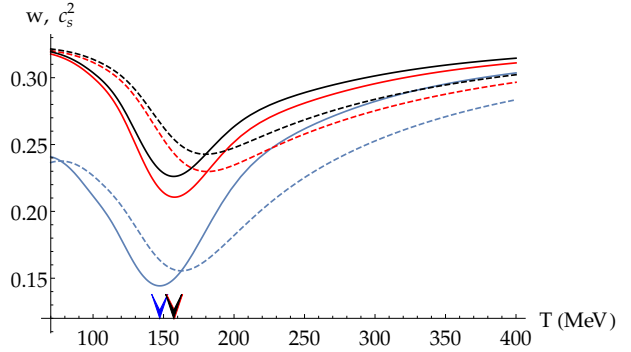}
	\caption{The speed of sound $c^2_s$ (continuous curves) and  $w$ (dashed line) for the different sectors: QCD (blue), QCD plus electroweak sector (red) and including a massless dark sector (black).} 
	\label{fig:cs}
\end{figure}

The results for the  $w=p_T/\varepsilon_T$  and for the speed of sound  $c^2_s$ are summarized in Fig.~\ref{fig:cs}, where the continuous curves indicate the speed of sound and the dashed lines the value of $w$. The blue lines give the results for the strong sector, the red ones contain the  electroweak sector and the black curves take into account the massless dark matter also. Cold dark matter is not included  since its contribution is  negligible.

The arrows indicate the temperature of the transition, defined as the temperature at the minimum of the speed of sound, which goes from the  $T_t^{s} = 147$~MeV for strongly interaction only, to $T_t^{ew} = 158$~MeV adding the electroweak sector, to $T_t^{d} = 157$~MeV including the massless dark sector.

\begin{figure}
	\centering
	\includegraphics[width=\columnwidth]{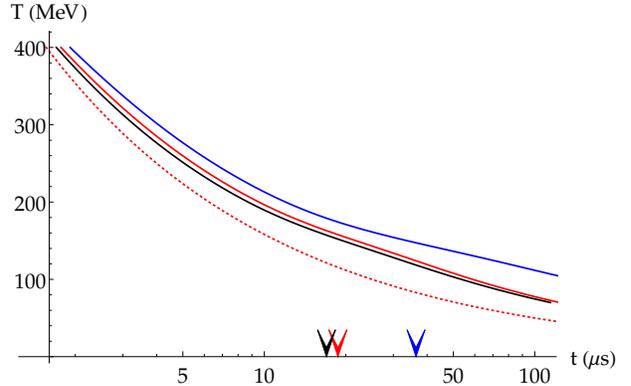}
	\caption{Temperature as a function of the cosmological time in the different sectors (see Fig.~\ref{fig:cs} for the colors legend), compared with the behavior of the pure radiation era (red dotted line).} 
	\label{fig:Tt}
\end{figure}

\begin{figure}
	\centering
	\includegraphics[width=\columnwidth]{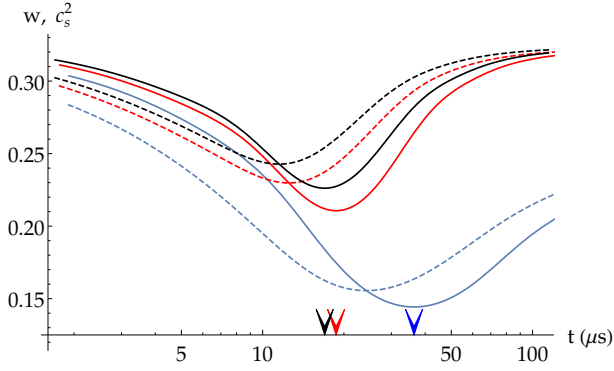}
	\caption{The sound speed $c^2_s$ (continuous line) and the EoS $w$ (dashed line) for the different sector (see Fig.~\ref{fig:cs} for the colors legend)	 as a function of time.} 
	\label{fig:csTIME}
\end{figure}

The relation between the temperature and the cosmological time is \begin{equation}
t = t_0 + \frac{1}{\sqrt{24\,\pi\,G}} \int_T^{T_0}\frac{d\overline T}{\overline T c^2_s\;\sqrt{\varepsilon}} \; ,
\end{equation} 
which is  numerically solved  (with $T_0 = 500$~MeV and $t_0=1\;\mu s$~\cite{Castorina2015}). In Figure~\ref{fig:Tt} we have shown how the temperature decrease in the different cases previously discussed and in the pure radiation era (red dotted line). The transition time is reduced by adding more and more sectors: $t_t^{s} =36.39 \;\mu s$, $t_t^{ew}= 18.71\;\mu s$ and $t_t^{d}=16.98\;\mu s$.

Finally, Figure~\ref{fig:csTIME} shows the behavior of the speed of sound  as a function of the cosmological time. For the whole system, after about $100 \mu s$ the values of $w$ and $c^2_s$ come back to be that ones of a radiation dominated era.

\section{Evolution of the cosmological parameters\label{sec:4}}

The results in the previous sections are the starting point to study the behavior of the cosmological parameters during the deconfinement transition. Since the cosmological parameters can depend on the higher order derivatives of the Hubble parameter, i.e. on the higher order derivative of the thermodynamical quantities, it could be possible  that some effects show up near the critical temperature~\cite{Castorina2015}.

We have analyze three different cases: strong sector only (blue curves in the figures); strong plus electroweak sector (red curves); the whole system, including the three EoS of section 2b for the dark matter (black lines). In all figures, the arrows indicate the transition time.

\begin{figure}
	\centering
	\includegraphics[width=\columnwidth]{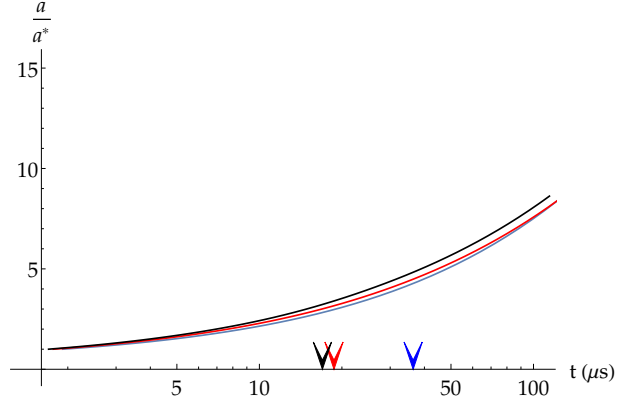}
	\caption{The scale factor $a/a^*$ as a function of cosmological time in the different sectors (see Fig.~\ref{fig:cs} for the colors legend).} 
	\label{fig:A}
\end{figure}

\begin{figure}
	\centering
	\includegraphics[width=\columnwidth]{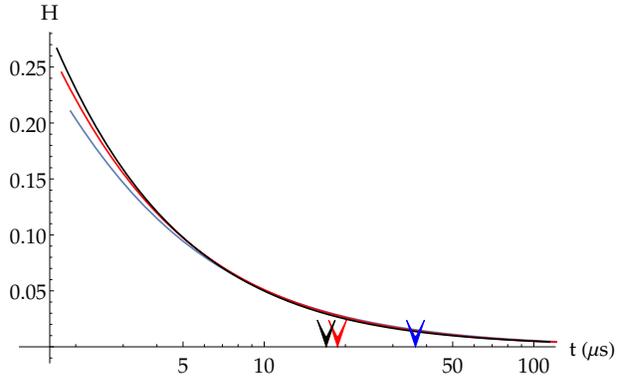}
	\caption{The Hubble parameter $H$ as a function of cosmological time in the different sectors (see Fig.~\ref{fig:cs} for the colors legend).} 
	\label{fig:H}
\end{figure}

In Figure~\ref{fig:A} and in Figure~\ref{fig:H} are respectively depicted the time behavior of the scale factor $a(t)$ (normalized to the value at $400$~MeV, $a^*$) and of $H(t)$. The final result is essentially independent on the specific setting.

\begin{figure}
	\centering
	\includegraphics[width=\columnwidth]{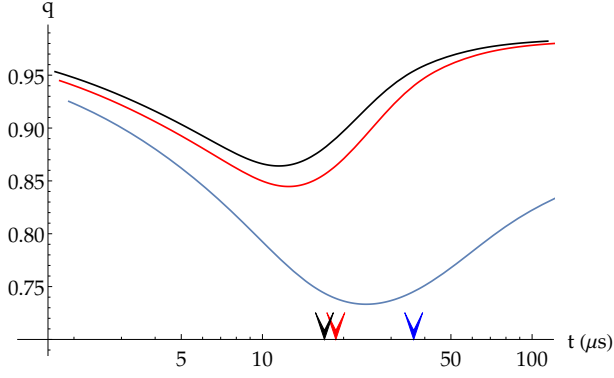}
	\caption{Cosmological deceleration $q$ as a function of cosmological time in the different sectors (see Fig.~\ref{fig:cs} for the colors legend).} 
	\label{fig:Qt}
\end{figure}

\begin{figure}
	\centering
	\includegraphics[width=\columnwidth]{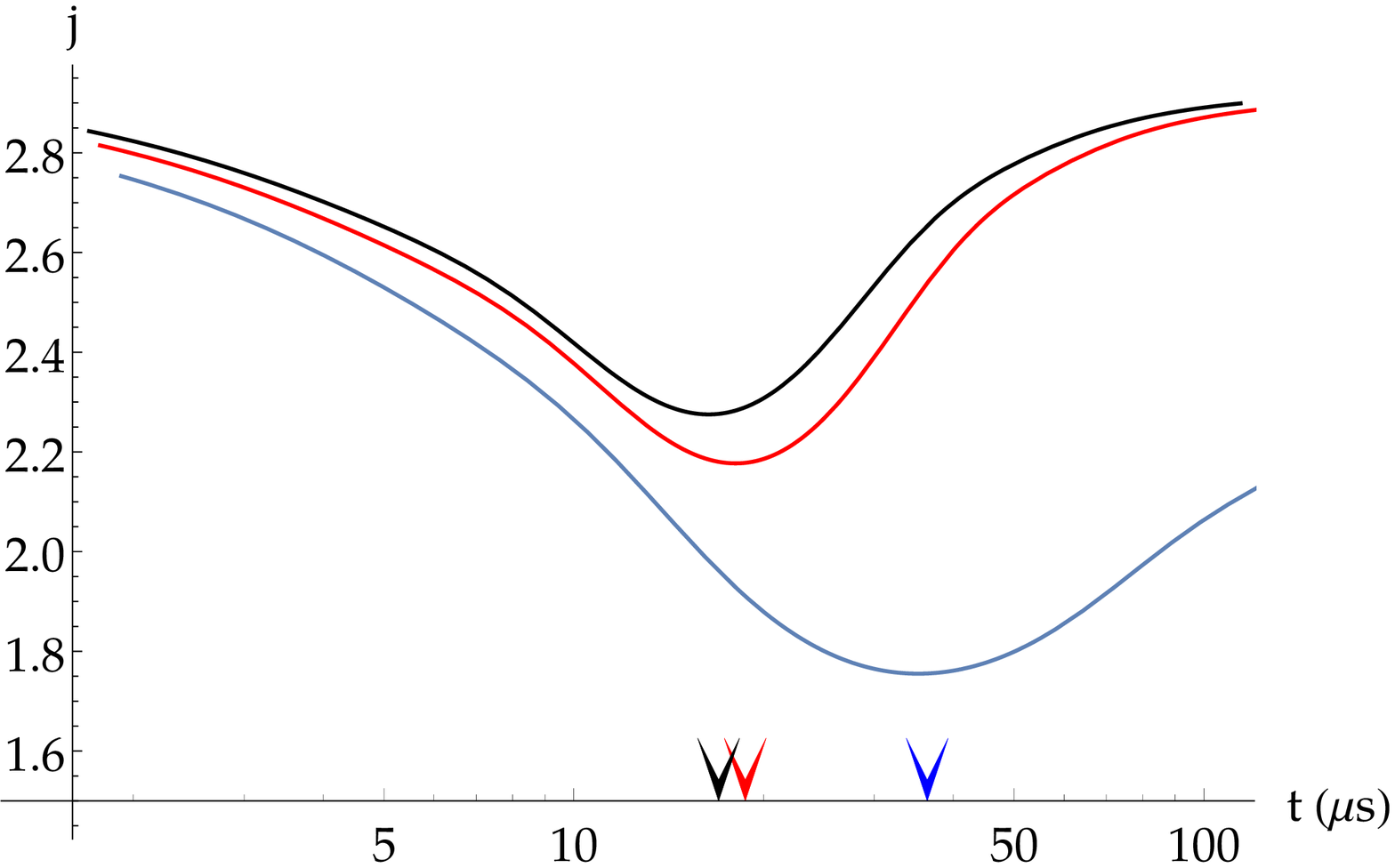}
	\caption{The jerk, $j$, as a function of cosmological time in the different sectors (see Fig.~\ref{fig:cs} for the colors legend).} 
	\label{fig:jt}
\end{figure}

\begin{figure}
	\centering
	\includegraphics[width=\columnwidth]{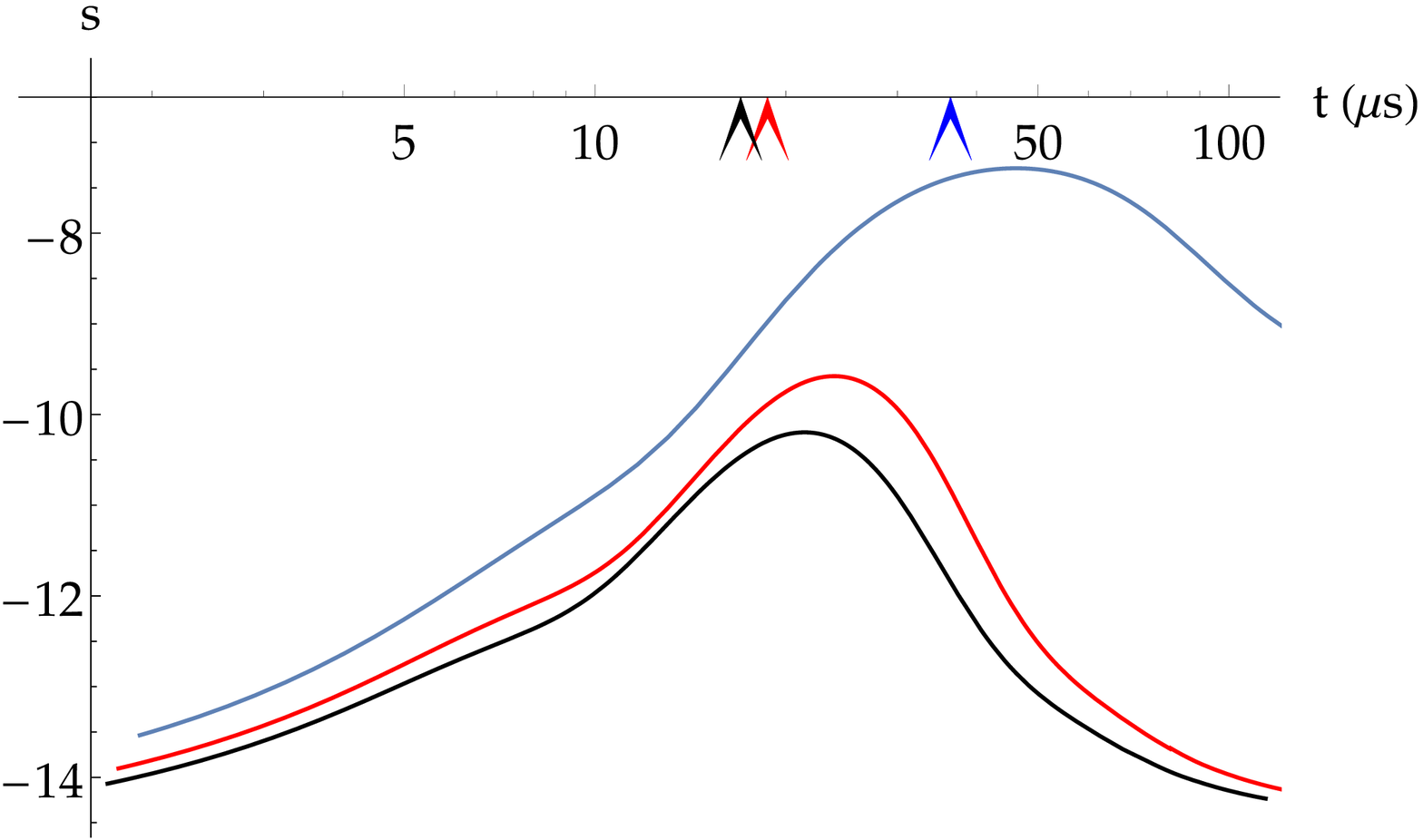}
	\caption{The snap, $s$, as a function of cosmological time in the different sectors (see Fig.~\ref{fig:cs} for the colors legend).} 
	\label{fig:A4t}
\end{figure}

\begin{figure}
	\centering
	\includegraphics[width=\columnwidth]{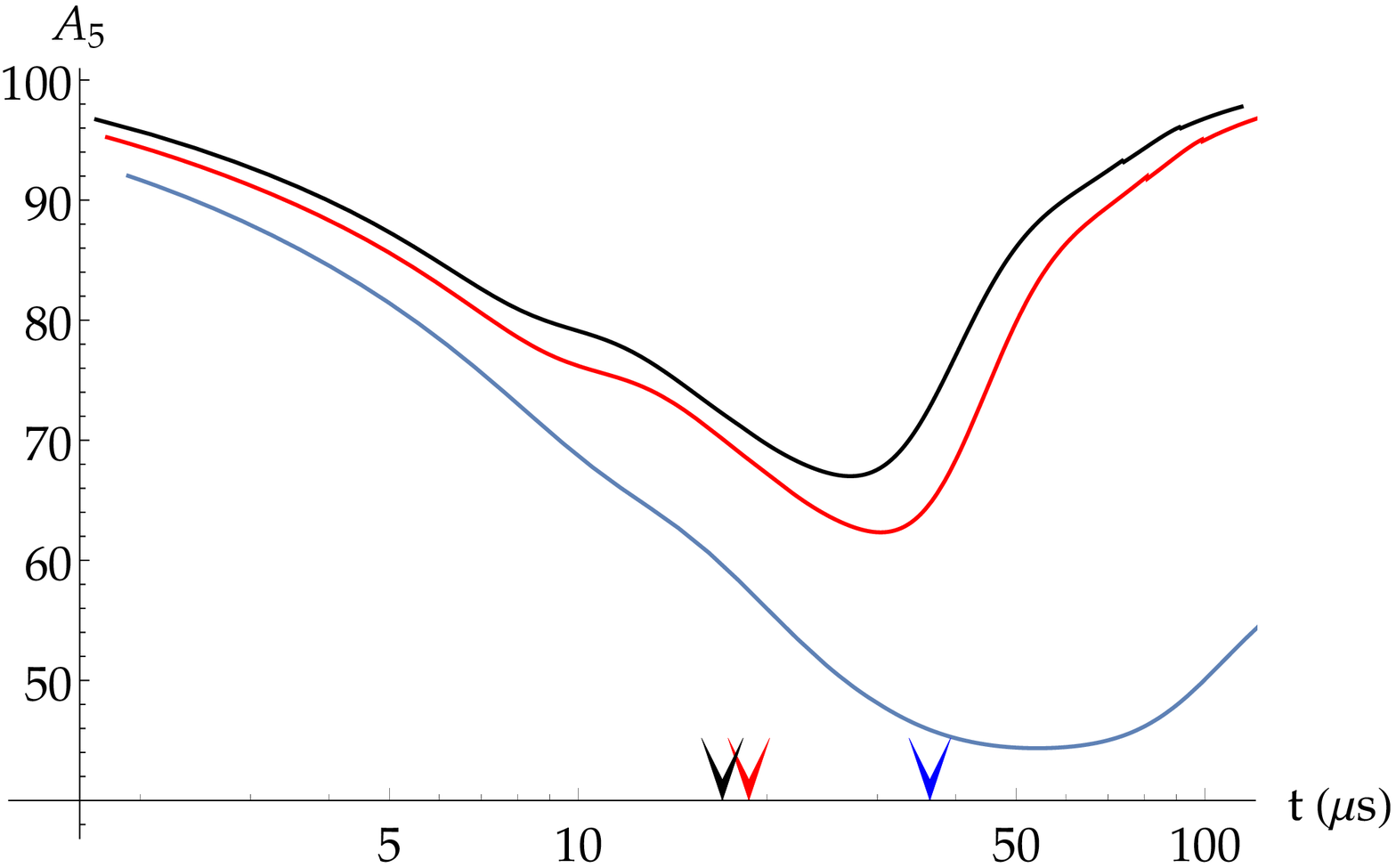}
	\caption{$A_5$ as a function of cosmological time in the different sectors (see Fig.~\ref{fig:cs} for the colors legend).} 
	\label{fig:A5t}	
\end{figure}

\begin{figure}
	\centering
	\includegraphics[width=\columnwidth]{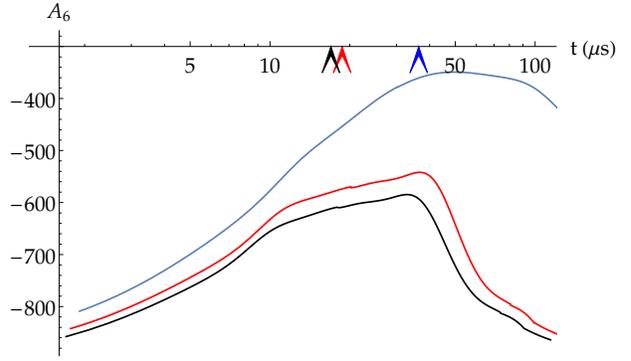}
	\caption{$A_6$ as a function of cosmological time in the different sectors (see Fig.~\ref{fig:cs} for the colors legend).} 
	\label{fig:A6t}
\end{figure}

In Figures~\ref{fig:Qt}, \ref{fig:jt}, \ref{fig:A4t}, \ref{fig:A5t} and \ref{fig:A6t} the time evolution of $q$, $j$, $s$, $A_5$ and $A_6$ is plotted. As expected the parameters with high order derivative show larger deviations from the typical values of  a radiation dominated era. However once the transition is over, the Universe is again dominated by radiation.

\vskip 120pt

\section{Modification of the primordial spectrum of the gravitational waves\label{sec:5}}


According to previous results, the fluctuations of the cosmological parameters in the whole system (strong,  electroweak and dark matter sectors)  are limited to a short time interval of about $100\;\mu s$. Therefore the deconfinement transition turns out to be more relevant in the modification of the primordial spectrum of the gravitational waves, proposed in~\cite{Schwarz1998}, that will be recalled in this section and reevaluated on the basis of the detailed description of the transition in Sec.~\ref{sec:3}.

During the inflation era the wavelengths of the quantum fluctuations are stretched to scales greater than the casually connected region and the fluctuations of the metric tensor result in a background of stochastic gravitational waves~\cite{Guzzetti2016}.

In the transverse traceless (TT) gauge, tensor perturbations $h_{ij}$ of the FLRW  metric satisfy the linearized equation of motion
\begin{equation}
{h_{ij;\mu}}^{;\mu} = 0 \; \label{eq:motionlin},
\end{equation} 
where ``;''  indicates the covariant derivative, and the corresponding  Fourier modes take the form
\begin{equation}
h_{ij}(\eta) = \int \frac{d^3k}{(2 \pi)^{3/2}} \, \sum_{\lambda} e_{ij}^{\lambda}h_{\bm{k},\lambda}(\eta)e^{i \bm{k \cdot x}} \; ,
\end{equation}
where $\lambda = (+, \, \times)$ are the two polarization states and $e_{ij}^{\lambda}$ is the symmetric polarization tensor ($e_{ii}=0$, $k^ie_{ij}=0$). In conformal time, $\eta$, the equation of motion for  the perturbations  reads~\cite{Watanabe2006}
\begin{equation}
h''_{\bm{k},\lambda}(\eta) + 2 \frac{a'}{a} h'_{\bm{k},\lambda}(\eta) + k^2h_{\bm{k},\lambda}(\eta) = 0 \; , \label{eq:motion1}
\end{equation}
where $d/d\eta$ is denoted by prime ``$^\prime$''. By defining 
$\mu_{\bm{k}\lambda} = a\;h_{\bm{k}\lambda} $,
Eq.~\eqref{eq:motion1} can be written as
\begin{equation}
\mu''_{\bm{k},\lambda}(\eta) +  \left( k^2 - \frac{a''}{a}\right) \mu_{\bm{k},\lambda}(\eta) = 0 \; \label{eq:motion2}.
\end{equation}
Two different regimes are physically relevant and  correspond to fluctuations well inside the Hubble horizon or well outside the horizon.
Since $a^{\prime\prime}/a \sim (a\;H)^2$, when $k \gg a\;H$ the wavelength is smaller than the horizon:  this is the \textit{subhorizon} regime. In this case Eq.~\eqref{eq:motion2} is the harmonic oscillator, hence, $\mu_{\bm{k}}(\eta)\sim e^{ik\eta}$ and for the perturbation one obtains
\begin{equation}
h_{\bm{k}} \sim a^{-1} \; ,
\end{equation}
which implies that  the amplitude decreases in time. In the \textit{superhorizon} regime, i.e. for $k \ll a\;H$, Eq. \eqref{eq:motion2} has two independent solutions: a decaying mode $\mu_{\bm{k}} \sim a^{-2}$, which we neglect, and $\mu_{\bm{k}} \sim a$  that leads to
\begin{equation}
h_{\bm{k}} \sim const. \; ,
\end{equation} 
that is the amplitudes are almost frozen, being outside the casually connected region. 

Therefore, during the inflation era the amplitudes are stretched to size larger than the horizon, where they remain constant, but when inflation ends the comoving Hubble horizon $(a\;H)^{-1}$ grows in time and each mode crosses the horizon and reenters inside the casually connected region when the wavelength is comparable to the horizon size, i.e. $k=a\;H$. 
In this case, a general solution of Eq.~\eqref{eq:motion2} can be written introducing a factor depending on mode's amplitude in superhorizon regime and a transfer function, $\mathcal{T}_k(\eta)$, as
\begin{equation}
h_{\bm{k},\lambda}(\eta) = h^{prim}_{\bm{k},\lambda}\mathcal{T}_k(\eta) \; ,\label{eq:solution}
\end{equation}
where $h^{prim}_{\bm{k},\lambda}$ is the amplitude when the mode left the horizon during the inflationary period and  $\mathcal{T}_k(\eta)$ describes the evolution of the gravitational wave after it crosses the horizon. In radiation dominated universe, the solution reads
\begin{equation}
h_{\bm{k},\lambda}(\eta) = h^{prim}_{\bm{k},\lambda} j_0(k \eta) \; , \label{eq:RadSolution}
\end{equation}
where $j_0(x)$ is the spherical Bessel function~\cite{Guzzetti2016}.
 
Let us define the power spectrum of gravitational waves. The energy density is given by 
\begin{equation}
\varepsilon_h(\eta) = \frac{1}{32\pi G a^2} <{{h'}_{ij}{h'}^{ij}}> \; . \label{eq:EnergyDensity}
\end{equation}
and in $k$ space the spatial average reads 
\begin{equation}
<{{h'}_{\bm{k},\lambda}{h'}_{\bm{k}',\lambda '}}> = (2\pi)^3 \delta_{\lambda \lambda'} \delta^{3}(\bm{k}+\bm{k}')|{h'}_{\bm{k},\lambda}|^2 \; .
\end{equation}
Moreover, one assumes that the primordial gravitational waves are unpolarized, that is $|{h'}_{\bm{k}, +}(\eta)|^2 = |{h'}_{\bm{k},\times}(\eta)|^2$.

Using Eq.~\eqref{eq:solution}, we can write the energy density as
\begin{equation}
\varepsilon_h(\eta) = \frac{1}{32 \pi G a^2} \int \frac{dk}{k} \Delta^2_{h, prim}\left[ \mathcal{T}'_k(\eta) \right]^2 \; ,
\end{equation}
where $\Delta^2_{k, prim} $ is the primordial amplitude which in de Sitter inflation turns out to be
\begin{equation}
\Delta^2_{h, prim} = \frac{2}{\pi^2} k^3 |h_{\bm{k}}^{prim}|^2 = \frac{16}{\pi} \left( \frac{H_{dS}}{M_{Pl}} \right)^2 \; , \label{eq:PrimAmplitude}
\end{equation} 
$H_{dS}$ and $M_{Pl}$ being the Hubble constant in de Sitter inflation and the Planck mass, respectively. 

The logarithmic energy density is defined as $d\varepsilon_h / d \ln k$ and  the fractional energy density is given by 
\begin{equation}
\Omega(\eta, k) = \frac{d \varepsilon_h(\eta, k)}{d \ln{k}} \frac{1}{\varepsilon_c(\eta)} = \frac{\Delta^2_{h, prim}\; \left[ \mathcal{T}'_k(\eta) \right]^2}{32 \pi\, G\, a^2\, \varepsilon_c(\eta)} \; ,\label{eq:EnDensity1}
\end{equation}
where $\varepsilon_c$ is the critical energy density.

From Friedman equations~\eqref{eq:FRWL} we finally get
\begin{equation}
\Omega(\eta, k) = \frac{\Delta^2_{h, prim}}{12\,H^2(\eta)\, a^2} \left[ \mathcal{T}'_k(\eta) \right]^2 \; . \label{eq:EnDensity2}
\end{equation}
A gravitational wave of mode $k$ has frequency $f = 2 \pi k/a$. Because of redshift, once a wave crosses the horizon its frequency decreases. From the definition of fractional energy density follows that $\Omega$ decreases as $a^{-4}H^{-2}$, since gravitational waves are decoupled from the rest of the Universe and $\varepsilon_c \sim H^2$ from Friedman equation. For waves that reentered at a certain time $\eta$, the fractional energy density today is 
\begin{equation}
\Omega_0 = \Omega(\eta, f) \frac{a^4(\eta) H^2(\eta)}{a^4_0 H^2_0} \; ,\label{eq:FreqToday}
\end{equation} 
and the frequency today is $f_0 = 2 \pi k / a_0$, where $a_0$ and $H_0$ are the scale factor and the Hubble parameter today.

The evolution of the $h_{\bm{k}}$ modes and of the crossing condition,  $k = a\;H$, are controlled by the scale factor $a$ and the modification of the spectrum of gravitational waves from the primordial one  depends on  the content of matter in the epoch they reenter the horizon. 
As previously shown (see Fig.~\ref{fig:csTIME}), during the QCD transition the speed of sound $c_s^2$ is strongly modified since the Universe stands no longer in a pure radiation era and, correspondingly, the primordial gravitational waves cross the horizon near that transition time at different rates. 

By lattice QCD simulations, the HRG model and including the electroweak and dark matter sectors, we now discuss a detailed analysis of this effect by numerical integration of Eqs.~\eqref{eq:motion1}-\eqref{eq:motion2}, improving previous analysis~\cite{Schwarz1998,Watanabe2006}.

It is more useful to write Eq.~\eqref{eq:motion1} for the transfer function as a function of the temperature, that is (see Appendix~\ref{app:c} for details)
\begin{equation}
\frac{d^2 \mathcal{T}_k }{d T^2} + f(T) \frac{d \mathcal{T}_k}{d T} + \kappa^2(T,k) \mathcal{T}_k= 0 \; .\label{eq:motion3}
\end{equation}
In order to integrate numerically Eq.~\eqref{eq:motion3}, we set boundary conditions  at high temperature, such as $10^4$~MeV, where the modes $h_{\bm{k}}$ are given by the radiation era solutions (Eq.~\eqref{eq:RadSolution}).

\begin{figure}
\centering 
\includegraphics[width=\columnwidth]{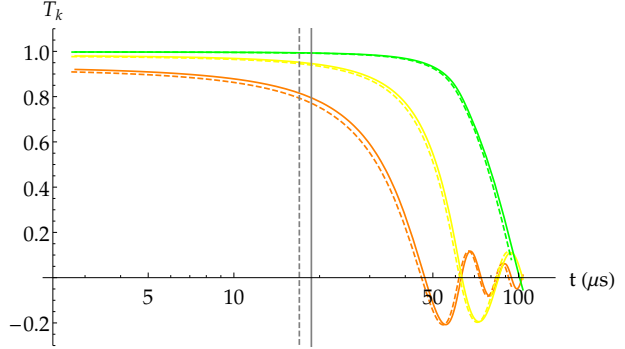} 
\caption{Transfer function $\mathcal{T}_k$ against cosmic time at different values of the wave number $k$. It describes the evolution of a gravitational wave. Green is for $k=2.17 \times 10^{-14} \; {\mu s}^{-1}$, yellow $k= 6.02 \times 10^{-14} \; {\mu s}^{-1}$, orange  $k = 1.20 \times 10^{-13 } \; {\mu s}^{-1}$. Continuous lines are for EW + QCD sectors, dashed lines contain massless Dark Matter. Vertical lines indicate the QCD transition.}{\label{fig:Solution}}
\end{figure}

In Figure~\ref{fig:Solution} the numerical results for different values of $k$ are reported.

Waves with higher frequencies cross the horizon earlier and waves that reentered at $T \sim 150$~MeV have frequencies of about $10^{-7}$~Hz,  the typical frequency $f_*$ of waves from the QCD transition.
The effects of the transition are expected to be impressed in the fractional energy density $\Omega$ and, in particular, one  computes~\cite{Schwarz1998}  the quantity $\Omega(f) / \Omega(\bar{f} \ll f_*)$, that is the fractional energy density of the gravitational waves with respect to the same quantity evaluated for waves that do not encounter the transition ($\bar{f}$ being a fixed frequency much lower than $f_*$).

\begin{figure}
\centering
\includegraphics[width=\columnwidth]{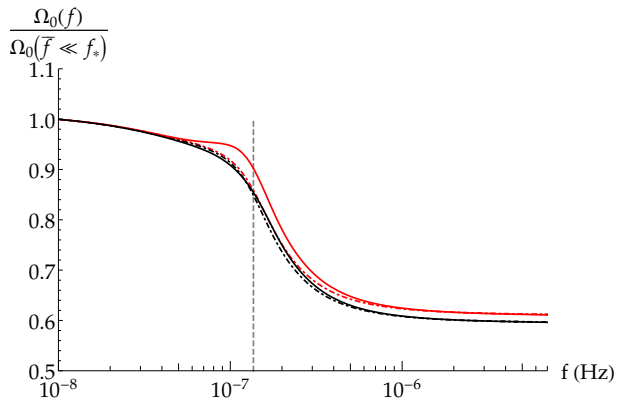} 
\caption{Fraction of energy density of gravitational waves with respect to waves that do not encounter the QCD transition in continuous lines, only the redshift factor to today values in dashed lines. Both against frequency $f$. EW contribution in red, massless Dark Matter in black. Vertical line represents the transition. The size of the step is about 38\%. } {\label{fig:Omega}} 
\end{figure}

\begin{figure}
\centering 
\includegraphics[width=\columnwidth]{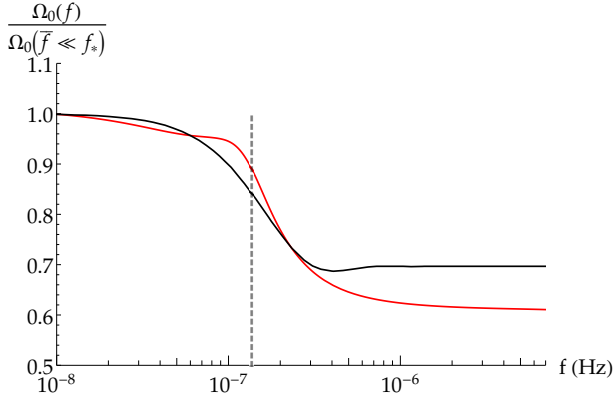} 
\caption{Comparison of the fraction of energy density of gravitational waves with respect to waves that do not encounter the QCD transition between the evaluation made in~\cite{Schwarz1998}, in black, and our evaluation, in red. Vertical line represents the transition.}{\label{fig:Compare}}
\end{figure}

From Eq.~\eqref{eq:FreqToday}, this quantity evaluated today is 
\begin{equation}
\frac{\Omega_0(f)}{\Omega_0(\bar{f} \ll f_*)} = \frac{\Omega(f)}{\Omega(\bar{f} \ll f_*)} \frac{a^4(f) H^2(f)}{a^4(\bar{f)}H^2(\bar{f})} \; . \label{eq:Omega} 
\end{equation}
The redshift factor gives the shape of the step and the final result is showed in Fig.~\ref{fig:Omega}. The size of the step is about 38\%, larger than previous results~\cite{Schwarz1998,Watanabe2006}. In particular, in~\cite{Schwarz1998} the step size was $\simeq 30\%$. Figure~\ref{fig:Compare} shows a direct comparison between the latter result and our evaluation. 

\label{fig}

\section{Comments and conclusions\label{sec:6}}

The fluctuations of conserved charges at the deconfinement transition are a clear signature  of the different behavior between a quark-gluon plasma and a hadron resonance gas model, but their detection in relativistic heavy ion collisions is difficult.

The fluctuations of the cosmological parameters at the QCD transition have, in principle, the same physical basis, i.e. they originate from the combined effect of the equation of state and of the calculation of higher order derivatives of the relevant physical parameters, that is, in early Universe, the scale factor.
 
We have shown, by a complete treatment of the thermodynamics of the whole system (strong, electroweak and dark matter contributions),  that after about $100 \mu s$ the cosmological parameters return to the typical values of a radiation dominated era, i.e. to their values before the transition, and that this result remains valid also for cosmological scalars involving higher order derivatives of the scale factor (see Figs. 6-10). 

Moreover the dark matter contribution turns out to be negligible, independently on the considered model and, as shown in ref. \cite{Greco2014}, the effects of the QCD transition on the density fluctuations are small.

Therefore the possible signature of the deconfinement transition in early Universe is restricted to the modification of the primordial gravitational wave spectrum. 

By using the recent lattice QCD simulation data and the HRG below $T_c$ to describe the transition, one evaluates the fraction of energy density of gravitational waves with respect to waves that do not encounter the QCD transition. A difference of about $10\%$ is observed with respect to previous analyses \cite{Schwarz1998}.
\vskip 40pt
{\bf Acknowledgement}

It is a pleasure to thank H.Satz and D.Schwarz for helpful comments.

\appendix

\section{Cosmological parameters \label{app:a}}
In the framework of Friedmann cosmology the evolution can be described by the Hubble parameter $H$, the deceleration $q$, the jerk $j=A_3$, the snap $s=A_4$ and others cosmological parameters, since they specify the various terms of the Taylor expansion of the scale factor:
\begin{equation}
\begin{split}
a(t) = a(t^*)\Big[ & 1 + H(t^*)\left(t-t^*\right)
- \\
& -
\frac{\left(q\,H^2\right)(t^*)}{2!}\left(t-t^*\right)^2 +\\
&
+
\frac{\left(j\,H^3\right)(t^*)}{3!}\left(t-t^*\right)^3+\cdots  
\Big]\;.
\end{split}
\end{equation}
Since
\begin{equation}
\frac{d^na}{dt^n} = \frac{d^{n-1}}{dt^{n-1}}\left(\frac{da}{dt}\right)  
=
\frac{d^{n-1}\left(a\,H\right)}{dt^{n-1}} \; ,
\end{equation}
it is easy to show that these quantities are related to the derivatives of the Hubble parameter as follows:
\begin{equation}
q =
-1 - \frac{\dot H}{H^2} \; ,
\end{equation}
\begin{equation}
j = A_3=
1 + 3\,\frac{\dot H}{H^2} + \frac{\ddot H}{H^3} \; ,
\end{equation}
\begin{equation}
s=A_4 = 
1 + 6\,\frac{\dot H}{H^2} + 3\,\left(\frac{\dot H}{H^2}\right)^2 + 4\,\frac{\ddot H}{H^3} + \frac{\dddot H}{H^4} \; ,
\end{equation}
\begin{equation}
\begin{split}
A_5 = & 1 + 10\,\frac{\dot H}{H^2} +
15\,\left(\frac{\dot H}{H^2}\right)^2+
10\,\frac{\ddot H}{H^3} + \\
& +
5\,\frac{\dddot H}{H^4} + 
10\,\frac{\dot H}{H^2}\,\frac{\ddot H}{H^3}+
\frac{H^{(4)}}{H^5} \; ,\\
\end{split} 
\end{equation}
\begin{equation}
\begin{split}
A_6 = & 
1 + 15\,\frac{\dot H}{H^2} + 45\,\left(\frac{\dot H}{H^2}\right)^2+
20\,\frac{\ddot H}{H^3} + \\
&
+ 15\,\frac{\dddot H}{H^4} + 
60\,\frac{\dot H}{H^2}\,\frac{\ddot H}{H^3}
+ 6\,\frac{H^{(4)}}{H^5} + \\
&
+15\,\left(\frac{\dot H}{H^2}\right)^3 + 
15\,\frac{\dot H}{H^2}\,\frac{\dddot H}{H^4} +\\
&
+10\,\left(\frac{\ddot H}{H^3}\right)^2 + 
\frac{H^{(5)}}{H^6}.
\end{split}
\end{equation}
By recalling that
\begin{equation}
w(\varepsilon)\equiv \frac{p}{\varepsilon},
\qquad 
c^2_s \equiv \frac{\partial p}{\partial \varepsilon} 
\end{equation}
and 
\begin{equation}
\frac{1}{H^{n+1}}\frac{d^nH}{dt^n} 
=
-\frac{4\,\pi\,G}{H^{n+1}}\;\frac{d^{n-1}\left(\varepsilon+p\right)}{dt^{n-1}}   \;,
\end{equation}
each of the previous derivatives can be express as
\begin{equation}\label{eq:Derivate H}
\frac{\dot H}{H^2} =
-\frac{3}{2}\left(1 + \frac{p}{\varepsilon} \right) \; ,
\end{equation}
\begin{equation}
\frac{\ddot H}{H^3} =  \frac{9}{2}\,\left(1 + c^2_s\right)\,\left(1 + \frac{p}{\varepsilon}\right) \; , \\
\end{equation}
\begin{equation}
\begin{split}
\frac{\dddot H}{H^4} = 
\frac{9}{2}\,\left(1 + \frac{p}{\varepsilon} \right)\Big[&
\frac{dc^2_s/dt}{H} - 3\,\left(1 + c^2_s\right)^2
-\\
&
-
\frac{3}{2}\,\left(1 + c^2_s\right)\left(1 + \frac{p}{\varepsilon} \right) \Big]  \; ,
\end{split} 
\end{equation}
\begin{equation}
\begin{split}
\frac{H^{(4)}}{H^5} =
\frac{9}{2}\,\Big(1 + &\frac{p}{\varepsilon}\Big)\Big[
9\left(1 + c^2_s\right)^3 
+ \\
&+ 18\,\left(1 + c^2_s\right)^2\,\left(1 + \frac{p}{\varepsilon} \right)
-\\
&
-3\,\left(4 + \frac{p}{\varepsilon}+3\,c^2_s \right)\,\frac{dc^2_s/dt}{H} +\\
&
+\frac{d^2c^2_s/dt^2}{H^2} \Big] \; ,
\end{split}
\end{equation}
\begin{equation}
\begin{split}
\frac{H^{(5)}}{H^6} &=
\frac{9}{2}\left(1 + \frac{p}{\varepsilon} \right)
\Big[
\frac{d^3c^2_s/dt^3}{H^3}  -
9 \left(\frac{dc^2_s/dt}{H}\right)^2 -\\
& - 
\left(\frac{33}{2} + \frac{9}{2}\,\frac{p}{\varepsilon}+12\,c^2_s\right)\frac{d^2c^2_s/dt^2}{H^2}    
+\\
&
+ \left(1+c^2_s\right)\left(135 + 81\,\frac{p}{\varepsilon} + 54\,c^2_s\right)\frac{dc^2_s/dt}{H} -\\
&
- 27\left(1+c^2_s\right)^4
- 
\frac{297}{2}\left(1+c^2_s\right)^3\left(1 + \frac{p}{\varepsilon}\right)
-\\
&-
27\left(1 + c^2_s\right)^2\left(1 + \frac{p}{\varepsilon}\right)^2
\Big] \; .
\end{split} 
\end{equation}
Furthermore, by the definition of $w(\varepsilon)$ one can show that
\begin{equation}
c^2_s = w+\varepsilon\;\frac{dw}{d\varepsilon} 
\end{equation}
and thus all the cosmological parameters can be express in terms of  $w$, $c^2_s$ and its derivatives. For the first three parameters one gets
\begin{equation}
q = \frac{1}{2}\left( 1 + 3\;w\left(\varepsilon\right)\right) \; ,
\end{equation}
\begin{equation}
\begin{split}
j = & 
1 + 3\,c^2_s\,\left(1 + q \right) 
=\\
= &
q\,\left(1+2\,q\right)
+3\,(1+q)\,\varepsilon\,\frac{dw}{d\varepsilon} \; ,
\end{split}
\end{equation}
\begin{equation}
\begin{split}
s = & 1 - 3 \,(1 + q) - 9\, c^4_s\, (1 + q) -\\ 
&-
3\, c^2_s (1+q)(3+q)
+
3(1 + q) \frac{dc^2_s/dt}{H}
=\\
=& - q\, (1+2\,q)\,(2 + 3\,q)
- \\
&-
3(1+q)(1+5\,q)\varepsilon\,\frac{dw}{d\varepsilon}
-\\
&- 
9 \,(1+q)\left(\varepsilon\,\frac{dw}{d\varepsilon}\right)^2 
+
3\, (1+q)\,\frac{dc^2_s/dt}{H} \; ,
\end{split}
\end{equation}
Finally, to simplify the calculations it is better to consider temperature derivatives rather than time derivatives. In general,  one needs a function $T=T(t)$ and, by defining the function 
\begin{equation}
h(T) = \frac{1}{a(T)}\,\frac{da}{dT}  \; ,
\end{equation}  
it is easy to show that
\begin{equation}
\frac{dT}{dt} = \frac{H}{h}  \; ,
\end{equation}
and, by the FLRW equations and by the isentropic expansion condition, one obtains
\begin{equation}
h = - \frac{1}{3\,c^2_s\,T} = - \frac{C_V}{3\,\left(\varepsilon+P\right)} \; ,
\end{equation}
where $C_V$ is the specific heat.

\section{Dark Matter Sector \label{app:b}}

\subsection{Cold Dark Matter \label{app:b1}} 
For a Cold Dark Matter one has
\begin{equation}\label{eq:cdmeos}
p_{cdm} = 0\;.
\end{equation}
and thus
\begin{equation}
H^2_{cdm} = \frac{8\,\pi\,G}{3}\;\varepsilon_{cdm} = \frac{M_{cdm}}{a^3}  \;.
\end{equation}
Moreover,  by the isentropic expansion condition, $s\,a^3=s_0\,a_0^3$, one obtains
\begin{equation}\label{eq:scdm} 
s = K\;\varepsilon_{cdm} \;,
\end{equation}
with 
\begin{equation}
K \equiv \frac{8\,\pi\,G}{3}\;\frac{s_0\,a_0^3}{M_{cdm}}   \;.
\end{equation}
If we indicate with an over-line the non-dark-matter contribution (for instance, the total entropy density is written as $s = \overline s + s_{cdm}
$), eq. \eqref{eq:scdm} gives
\begin{equation}
s = \overline s +s_{cdm} = K\,\varepsilon_{cdm}\:.
\end{equation}
Since for cold dark matter $T\,s_{cdm} = \varepsilon_{cdm}$, one finds  
\begin{equation}\label{eq:cdm}
\varepsilon_{cdm}
=
\frac{\overline s\,T}{K\,T-1} =  
\frac{\overline \varepsilon + \overline  p}{K\,T-1}\;.
\end{equation}
The value of $K$ can be obtained by specifying  eq. \eqref{eq:cdm} at present time
\begin{equation}
\begin{split}
K =& \left. \frac{1}{T}\left(1 + \frac{\overline \varepsilon + \overline p}{\varepsilon_{cdm}}\right)\right|_{T=T_0} =\\
=&  \frac{1}{T_0}\left(1 + \frac{\Omega_{b0}}{\Omega_{cdm0}}  \right)  \sim 5\times 10^9 \, MeV^{-1} \;,
\end{split}
\end{equation}
with $\Omega_{b0}$ and $\Omega_{cdm0}$ the present baryon and cold dark matter densities respectively.

Since $K\,T\gg 1$, $\varepsilon_{cdm}\sim 10^{-10}\;\overline \varepsilon$ and the cold dark matter contribution is negligible. 

\subsection{General Barotropic case \label{app:b2}}
A barotropic fluid is characterized by a linear equation of state
\begin{equation}\label{eq:beos}
p_{b} = w_b\; \varepsilon_b \;,
\end{equation}
with $w_b = const.$ and by energy conservation one finds that
\begin{equation}
H^2_{b} = \frac{M_b}{a^{3(1+w_b)}} \;.
\end{equation}
By using the isentropic expansion condition and recalling that for  the barotropic entropy density, $s_b$, one has $T\,s_b = (1 + w_d)\,\varepsilon_d$, it turns out that
\begin{equation}\label{eq:difsbdm}
\overline s = 
K_{w_b} \left(
\frac{T}{T^{\star^{\scriptstyle 4}}}
\; \frac{s_b}{1 + w_b}\right)^{\textstyle{\frac{1}{1+w_b}}}
-s_b \; ,
\end{equation}
where 
\begin{equation}
\begin{split}
K_{w_b} &\equiv  s_0\;a_0^3\;\left(\frac{8\,\pi\,G}{3}\;\frac{T^{\star^{\scriptstyle 4}}}{M_b}\right)^{\textstyle{\frac{1}{1+w_b}}} 
=\\
&= 
\left. s\;\left(\frac{T^{\star^{\scriptstyle 4}}}{\varepsilon_b}\right)^{\textstyle{\frac{1}{1+w_b}}}\right|_{T=T_0}
\end{split} 
\end{equation}
and we have defined a conveniently temperature $T^\star$ to make the dimension of $K_{w_b}$ independent to $w_b$.

Eq.~\eqref{eq:difsbdm} can not be inverted analytically for all $w_b$, but only for a few values of $w_b$ ($w_b=0$ gives the cold dark matter case).

\subsection{Polytropic case and Chaplygin gas \label{app:b3}} 
A polytropic fluid is characterized by the equation of state  
\begin{equation}
p_p = p^0_p\;\left(\frac{\varepsilon_p}{\varepsilon^0_p} 
\right)^{\gamma} =
w^0_p\;\varepsilon^0_p\;\widehat{\varepsilon}_p^\gamma \;,
\end{equation}
where $p^0_p$, $\varepsilon^0_p$ are reference values, $w^0_p \equiv p^0_p/\varepsilon^0_p$ and $\widehat \varepsilon_p\equiv\varepsilon_p/\varepsilon^0_p$.
 
In this case the equation of energy conservation becomes
\begin{equation}
\frac{d\widehat \varepsilon_p}{dt} = -3\,\left(\frac{1}{a}\,\frac{da}{dt} \right)\,\left(\widehat \varepsilon_p + w^0_p\;\widehat \varepsilon^{\gamma}_p \right) \;.
\end{equation}
The integration from $a_0=a(\varepsilon^0_p)$ to $a$ gives
\begin{equation}
\widehat \varepsilon_p = 
\left[-w^0_p + \frac{w^0_p+1}{\widehat a^{3(1-\gamma)} } \right]^{\textstyle{\frac{1}{1-\gamma}}} \; ,
\end{equation}
with $\widehat{a}\equiv a/a_0$. Thus, in a cosmological context, a polytropic fluid corresponds to a generalized Chaplygin gas with index $\alpha=-\gamma$. 

\section{Calculation of the  temperature dependence \label{app:c}}
Let us consider the equation of motion for the transfer function with respect to cosmic time $t$
\begin{equation}
\frac{d^2 \mathcal{T}_k}{d t^2} + 3 \frac{1}{a}\frac{d a}{d t} \frac{\mathcal{T}_k}{d t}+ \frac{k^2}{a^2}\mathcal{T}_k = 0 \label{eq:motionApp}
\end{equation}
and let us write 
\begin{equation}
\frac{d T}{d t} = \frac{d T}{d \eta} \frac{d \eta}{d t} = - 3\, c_s^2\, T\, H \; .
\end{equation}
Then Eq.~\eqref{eq:motionApp} becomes
\begin{equation}
\frac{d^2 \mathcal{T}_k }{d T^2} + f(T) \frac{d \mathcal{T}_k}{d T} + \kappa^2(T,k) \mathcal{T}_k= 0 \; ,
\end{equation}
where 
\begin{align}
f(T) &= \frac{1}{T} \frac{w - 1 + 2 c_s^2}{2 c_s^2} + \frac{1}{c_s^2}\frac{d c_s^2}{d T} \; , \\
\kappa(T,k) &= -\frac{k}{a}\frac{1}{3 c_s^2 T H} \; , \\
w &= \frac{P}{\varepsilon} \; , \\
c_s^2 &= \frac{dP}{d\varepsilon} \; .
\end{align}
In radiation era, the solution  of Eq.~\eqref{eq:motionApp} reads
\begin{equation}
\mathcal{T}_k = A \,  j_0\left(\alpha\; \frac{k}{T} \right) \; ,
\end{equation}
where $A$ and $\alpha$ are appropriate constants.

\addcontentsline{toc}{section}{References}

\end{document}